\documentclass[3p]{elsarticle}

\usepackage{lineno,hyperref}
\modulolinenumbers[5]
\usepackage{amsmath}
\usepackage{amssymb}
\usepackage{booktabs,multirow}

\newcommand\ba{\begin{eqnarray}}
\newcommand\ea{\end{eqnarray}}

\makeatletter
\def\ps@pprintTitle{%
   \let\@oddhead\@empty
   \let\@evenhead\@empty
   \def\@oddfoot{\reset@font\hfil\thepage\hfil}
   \let\@evenfoot\@oddfoot
}
\makeatother

\begin{document}

\begin{frontmatter}

\title{Arbitrary $\ell$-state solutions of the Klein-Gordon equation with the Manning-Rosen plus a Class of Yukawa potentials}

\author[bakuaddress,bakuaddress2]{A.~I.~Ahmadov}
\ead{ahmadovazar@yahoo.com}

\author[ktuaddress]{M. Demirci\corref{mycorrespondingauthor}}
\cortext[mycorrespondingauthor]{Corresponding author}
\ead{mehmetdemirci@ktu.edu.tr}

\author[bakuaddress]{S. M.~Aslanova}
\ead{sariyya.aslanova@mail.ru}

\author[ktuaddress]{M. F. Mustamin}
\ead{mfmustamin@ktu.edu.tr}

\address[bakuaddress]{Department of Theoretical  Physics, Baku State
University,\\ Z. Khalilov st. 23, AZ-1148, Baku, Azerbaijan}
\address[bakuaddress2]{Institute  for Physical Problems, Baku State
University,\\ Z. Khalilov st. 23, AZ-1148, Baku, Azerbaijan}
\address[ktuaddress]{Department of Physics, Karadeniz Technical
University, TR61080 Trabzon, Turkey}

\begin{abstract}
Focusing on an improved approximation scheme, we present how to treat the centrifugal and the Coulombic behavior terms and then to obtain the bound state solutions of the Klein-Gordon (KG) equation with the Manning-Rosen plus a Class of Yukawa potentials. By means of the Nikiforov-Uvarov (NU) and supersymmetric quantum mechanics (SUSYQM) methods, we present the energy spectrum for any $\ell$-state and the corresponding radial wave functions  in terms of the hypergeometric functions. From both methods we obtain the same results. Several special cases for the potentials which are useful for other physical systems are also discussed. These are consistent with those results in previous works. We obtain that the energy level $E$ is sensitive to the potential parameter $\delta$ at fixed values of other parameters and increases when $\delta$ runs from $0.05$ to $0.3$. Furthermore, $E$ is sensitive to the quantum numbers $\ell$ and $n_r$ for a given $\delta$, as expected.
\end{abstract}


\end{frontmatter}
%

\section{Introduction}
Molecules, atoms, nuclei, etc., in order to obtain experimental information on their structures and interactions, are bombarded with the beams of high-energy particles. These are known as scattering experiments. On the other
hand, theoretical researches are carried out by examining the non-relativistic or the relativistic wave equations for
any given potential. In quantum mechanics (QM), an analytical solution in the form of a wave function is required since this form contains all the important properties for a quantum system to be definable properly~\cite{Greiner, Bagrov, Gara, Boivin,
Iwo, Mili}. Moreover, the particle's dynamics in high energy can be described by the prescription of
the relativistic wave equations as in the subject of particle and nuclear physics~\cite{Greiner, Bagrov, Herman}.
For the case of scalar particles, the particle motion
obeys the Klein-Gordon (KG) equation~\cite{KFG1,KFG2,KFG3,KFG4}. Therefore, the KG equation analytical
solutions with interaction potentials play a significant role in relativistic QM.
Notice that for the case interaction potential is not sufficient to create particle and anti-particle pairs, the KG equation can be applied to treat spin-$0$ particle as for the Dirac equation describes spin-$1/2$ particle.
Placing particle in a condition with strong potential field, the quantum system is in a relativistic effect and hence gives correction to the non-relativistic case.

There are many techniques to solve the wave equations with potentials in the relativistic
and also non-relativistic circumstances. The following are some of them: Nikiforov-Uvarov (NU) method~\cite{Nikiforov}, supersymmetry QM (SUSYQM) \cite{Cooper1,Cooper2,Morales}, shifted 1/N expansion method \cite{Tang,Roy}, asymptotic iteration method~\cite{aim}, Hartree-Fock method~\cite{Hartree}, the path integral method~\cite{Cai}, factorization~\cite{Dong1} and perturbation theory~\cite{Stevenson}. Among them,
the NU and SUSYQM methods have received great interest. By using these two techniques, many works have been conducted to obtain either exact or approximate solutions of the KG equation with some well-known potentials as follows: Manning-Rosen Potential~\cite{Dong3,Jia2013,Wei2010}, Yukawa potential~\cite{Sever2011,Hamzavi2013,Wang2015}, Hulthen Potential~\cite{Znojil,Yuan,Ikot2011}, generalized Hulthen potential~\cite{Mehmet,Sever,Qiang}, Kratzer Potential~\cite{Qiang2004}, Wood-Saxon Potential~\cite{Guo2005,Berkdemir,Badalov} and Deng-Fan molecular potentials~\cite{Oluwadre}. Similarly for the case of combined potentials: Manning–Rosen plus Hulthén potential~\cite{Ahmadov1}, Hulthén plus a Ring-Shaped like potential~\cite{Ahmadov2}, Hulthén plus Yukawa potential~\cite{Ahmadov3} and references in there~\cite{Ahmadov4}. Particularly,  most of them based on the solutions of the KG equation with equal and unequal vector and scalar potential energies.

Although those previous attempts have provided satisfactory bound state solutions of the KG equation by using Manning-Rosen and Yukawa potentials separately, no one considers the KG equation under their linear combination for an arbitrary $\ell$ state, so far. Therefore, in this study we examine the bound state solutions for this combined potential in the framework of the KG equation.
The Manning-Rosen potential can be utilized to represent an interaction system that contains the continuum and bound-states, and then applied to various research fields such as atomic, condensed matter, particle and nuclear physics. For a particle under this potential, the relativistic effects can become significant, especially for strong coupling. The Manning-Rosen potential is defined by~\cite{Manning1,Manning2}
\ba
V_{MR}(r)=\frac{\hbar ^{2}}{2M b^{2}}\left[\frac{\eta(\eta-1)e^{-2r/b}}{(1-e^{-r/b})^2}-\frac{A
e^{-r/b}}{(1-e^{-r/b})}\right], \label{a1} \ea
in which the parameter $b$ relates to the potential range while $A$ and $\eta$ are two dimensionless parameters.
This kind of potential is used to describe the vibrations of diatomic molecules
and in addition, forms an appropriate model for other physical events.

On the other hand, the Yukawa potential~\cite{Yukawa} is an effective potential in a non-relativistic
realm which describes the nucleon strong interactions. It is defined as
\ba
V_{Y}(r)=-\frac{V_{0} e^{-\delta r}}{r} \label{a2}
\ea
where $V_{0}$ is the strength of the potential and $1/\delta$ is its range. This potential is also known as the
Debye-H\"{u}ckel potential in plasma physics, where it describes a charged particle in a weakly non-ideal plasma, as
well as in colloids and electrolytes. However, in this study we consider a Class of Yukawa potential defined by
\ba
V_{CY}(r)=-\frac{V_{0} e^{-\delta r}}{r}-\frac{V'_{0} e^{-2\delta r}}{r^2}. \label{potCY}
\ea

Briefly, both potentials are two screened Coulomb potential in simple representation, i.e., in small $r$ they have a Coulombic behavior but then descend exponentially as $r$ becomes larger. Their linear combination can be utilized to study the nucleus deformed-pair
interactions and spin-orbit coupling in the potential field. The additional charming
viewpoint of this potential is that it may be used to describe the vibration of the hadronic
system and can also be formed for a convenient model in other physical phenomena. From the investigation of the KG equation under the linear combination potential, one can provide the deeper and accurate appreciation of the physical properties of the wave functions and the energies in the continuum and  bound states of the interacting systems. Based on all the backgrounds
and previous works, in this study we focus on the following linear combination of Manning-Rosen and a Class of Yukawa potentials:
\ba
V(r)=\frac{\hbar ^{2}}{2M b^{2}}\left[\frac{\eta(\eta-1)e^{-2r/b}}{(1-e^{-r/b})^2}-\frac{A
e^{-r/b}}{(1-e^{-r/b})}\right]-\frac{V_{0} e^{-\delta r}}{r}-\frac{V'_{0} e^{-2\delta r}}{r^2}.\label{a4}\ea
Our objective is to study this potential in a subsequently large quantum system. For this aim, we apply NU and SUSYQM~\cite{Gendenshtein1,Gendenshtein2} methods to the problem, and use a scheme of improved approximation to handle the centrifugal and Coulombic behavioral terms. As the results, we obtain the energy eigenvalues and the normalized radial wave functions for any $\ell$. The same problem for $\ell=0$ have been studied in Ref.~\cite{Ita,Louis} as well, but our outcome disagrees with the result therein.

We arrange this paper as follows:  In Sec.~\ref{br}, we introduce the KG equation with the Manning-Rosen plus a Class of Yukawa potentials under an improved approximation scheme. In Sec.~\ref{bss}, the bound state solutions of the KG equation are obtained by using the NU (Sec.~\ref{NUmethod}) and SUSYQM (Sec.~\ref{sr}) methods, separately. The particular cases are discussed in Sec.~\ref{pc}. Next, in Sec.~\ref{nr}, we present the numerical results for the energy levels depending on potential parameters $\delta$ and quantum numbers $n,\ell$. Finally, Section~\ref{cr} provides the concluding remarks of our work.

\section{Governing Equation}\label{br}

Two types of potential coupling can be introduced into the KG equation
since it consists of two objects: the operator of 4-vector linear
momentum $P_\mu$ and the scalar rest mass $M$. The first
type is a scalar potential ($S$)(via the substitution $M \to M + S$) and the second is a vector potential (via minimal coupling $P_\mu \to P_\mu-g A_\mu$)~\cite{Greiner}. Gauge invariance of the vector coupling provides the freedom to fix the gauge without changing the physical meaning of the problem. Consequently,
it is possible that the potentials with two types coupling are the space-time $S$-potential and the four $V$-potential as $g A_0 = V(t,r)$. We consider the time-independent KG equation for a time-independent $V$ and $S$ potentials and
in a region without a magnetic field but a vector potential as follows:
\ba
\nabla^{2}\psi+\frac{1}{(\hbar c)^2}\biggl[(E-V)^2-(M c^2+S)^2 \biggr]\psi=0,
\label{KGequ}
\ea
where $E$ denotes the system relativistic energy. In the natural units, $\hbar = c = 1$, this equation is written as
\ba
[-\nabla^{2}+(M+S(r))^{2}]\psi(r,\theta,\phi)=[E-V(r)]^{2}\psi(r,\theta,\phi).
\label{a5}
\ea
In the framework of the spherical coordinates system, the wave function $\psi(r,\theta,\phi)$, which is a solution of the above equation, can be divided into radial and angular dependencies as follows:
\ba
\psi(r,\theta,\phi)=\frac {\chi(r)}{r}\Theta
(\theta)e^{im\phi},~~m \in \mathbb{Z} =0,\pm 1,\pm 2,\ldots\label{a6}
\ea
Placing Eq.\eqref{a6} into Eq.\eqref{a5} gives the radial differential equation as follows:
\ba
\begin{split}
\chi^{''}(r)+&\biggl[(E^{2}-M^{2})-2( S(r)\cdot M+ V(r)\cdot E)+(V^{2}(r)-S^{2}(r))-\frac{\ell(\ell+1)}{r^{2}}\biggl] \chi(r)=0.
\end{split}
\label{a7}
\ea

In this study, we regard that the vector potential is equal to the scalar
potential and this leads to the following equation:
\ba
\chi^{''}(r)+\biggl[(E^{2}-M^{2})-2(E+M) V(r)
-\frac{\ell(\ell+1)}{r^{2}}\biggr] \chi(r)=0. \label{a8}
\ea

The above relation with the combined potential~\eqref{a4}
can be exactly solved for $\ell\neq0$, only if it is made an approximation when we deal with the centrifugal and the Coulombic behavior terms. By using the approximation scheme proposed by Greene and Aldrich~\cite{Greene}, the centrifugal term can be approximately expressed by~\cite{Wen1,Wei,Dong6},
\ba \frac {1}{r^{2}}\approx
{4\delta^{2}}\frac{e^{-2\delta r}}{(1-e^{-2\delta r})^2},
\label{a10}
\ea
from which we have
\ba
\frac{1}{r}\approx{2\delta}\left[\frac{e^{-\delta r}}{(1-e^{-2\delta
r})}\right]. \label{a9}
\ea
This approximation is valid in the case of $\delta r < < 1$.

We now rewrite the Manning-Rosen potential under assumption of $1/b=2\delta$ as follows:
\begin{equation} \label{a11}
\begin{split}
V'_{MR}(r)&=\frac{\hbar^2}{2M b^2}\left[\frac{\eta(\eta-1)e^{-2r/b}}{(1-e^{-r/b})^2}-\frac{Ae^{-r/b}}{(1-e^{-r/b})}\right]\\
&=\frac{V_{01}e^{-4\delta r}}{(1-e^{-2\delta r})^2}- \frac{V_{02}e^{-2\delta r}}{1-e^{-2\delta r}},
\end{split}
\end{equation}
where
\ba V_{01}= \frac{2\hbar^2 \delta^2 \eta (\eta-1)}{M},
\,\,\,\ V_{02}= \frac{2\hbar^2 \delta^2 A}{M}. \label{a11}
\ea

If the approximation is applied to the Class of Yukawa potential~\eqref{potCY}, then it reads:
\ba V'_{CY}(r)=-\frac{2 \delta  V_0 e^{-2\delta r}}{1-e^{-2\delta r}}-\frac{4 \delta^2  V'_0 e^{-4\delta r}}{(1-e^{-2\delta r})^2}=
-\frac{V_{03} e^{-2\delta r}}{1-e^{-2\delta r}}+\frac{V_{04} e^{-4\delta r}}{(1-e^{-2\delta r})^2} \label{a12} \ea
where
\ba V_{03}= 2\delta V_0, V_{04}=- 4 \delta^2  V'_0. \label{a13}
\ea

Therefore, after application of approximation scheme, a linear combination of Manning-Rosen and a Class of Yukawa potentials becomes
\ba
\begin{split}
V'(r)&=V'_{MR}(r)+V'_{CY}(r)\\
&=\frac{(V_{01}+V_{04})e^{-4\delta
r}}{(1-e^{-2\delta r})^2}- \frac{(V_{02}+V_{03})e^{-2\delta
r}}{1-e^{-2\delta r}}= \frac{V_{014}e^{-4\delta r}}{(1-e^{-2\delta
r})^2}-\frac{V_{023}e^{-2\delta r}}{1-e^{-2\delta r}}\label{a14}
\end{split}
\ea
where $V_{014}=(V_{01}+V_{04})$ and $V_{023}=(V_{02}+V_{03})$.
\begin{figure}[bh]
    \begin{center}
\includegraphics[scale=0.42]{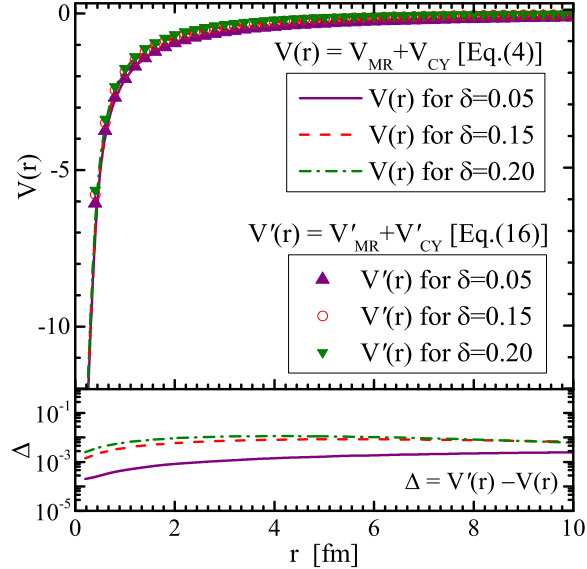}
     \end{center}
 \vspace{-4mm}
\caption{The variations of total potential and its approximation~(\ref{a14}) with respect to separation distance $r$ for some values of parameter $\delta$. Here, we take the some parameters as $V_0=1, V'_0=0.1, \eta=0.75$, M=1 and $A=2b$.}
\label{fig1:potapp}
\end{figure}
To have a quantitative understanding of the effect of approximation
on potential, the total potential~\eqref{a4}, its approximation~\eqref{a14} and their difference are given as a function of $r$ for different values of $\delta$  in Fig.~\ref{fig1:potapp}. It is clear that the approximation becomes more suitable for small values of $\delta$. The difference $\Delta$ is about at order of $10^{-3}$, depending on potential parameters. It means that the equation~\eqref{a9} is a good approximation for centrifugal term as the parameter $\delta$ becomes small.

Furthermore, we might use the approximation~\eqref{a10} to the centrifugal term in Eq.~\eqref{a8}. As a result, the equation~\eqref{a8} is expressed in the following form
\ba
\begin{split}
\chi^{''}(r)+\biggl[(E^{2}-M^{2})-2(M+E)\left(\frac{V_{014}e^{-4\delta
r}}{(1-e^{-2\delta
r})^2}-\frac{V_{023}e^{-2\delta r}}{1-e^{-2\delta r}}\right)\\
-\frac{4\ell(\ell+1)\delta^{2}e^{-2\delta
r}}{(1-e^{-2\delta r})^2}\biggr] \chi(r)=0, \label{a15}
\end{split}
\ea
under the considered approximation. Here, the effective potential is defined by
\ba
V_{\rm eff}(r)= 2(E+M)\left[\frac{V_{014}e^{-4\delta
r}}{(1-e^{-2\delta r})^2}- \frac{V_{023}e^{-2\delta
r}}{1-e^{-2\delta r}}\right]+\frac{4\ell(\ell+1)\delta^{2}e^{-2\delta
r}}{(1-e^{-2\delta r})^2}.\label{a16}
\ea

\section{Bound State Solutions for the Manning-Rosen plus a Class of Yukawa potentials}\label{bss}
In this section, we discuss how to obtain the bound state solution of the KG equation radial dependency by implementing the NU and SUSYQM methods, respectively.
\subsection{Implementation of NU Method}\label{NUmethod}
For implementing the NU method, the differential equation~\eqref{a15} must be transformed to the following hypergeometric type equation form:
\ba
\chi^{''}(s)+\frac{\tilde{\tau}}{\sigma}\chi^{'}(s)+\frac{\tilde{\sigma}}{\sigma^2}\chi(s)=0.
\label{a17}
\ea

The solutions of this equation must satisfy $\chi(0)=0$ and $\chi(\infty) \rightarrow 0$ boundary conditions. Applying the transformation $s=e^{-2\delta r}\in[0,1]$ for $r \in[0,\infty)$, the equation~\eqref{a15} takes the following form:
\ba
\begin{split}
\chi''(s)+\frac{1-s}{s(1-s)}\chi'(s)+\biggl[\frac{1}{s(1-s)}\biggr]^2&\biggl[-\varepsilon^2(1-s)^2-\alpha^{2}s^2 \\
&+\beta^{2}s(1-s)-s\ell(\ell+1)\biggr]
\chi(s)=0,\label{a20}
\end{split}
\ea
with
\ba
\varepsilon=\frac{\sqrt{M^2-E^2}}{2\delta}>0~,~~\alpha
=\frac{\sqrt{2V_{014}(E+M)}}{2\delta}>0~,~~\beta
=\frac{\sqrt{2V_{023}(E+M)}}{2\delta}>0, \label{a19}
\ea
where $E$ must be smaller than $M$, i.e.,  $E<M$. The equation~\eqref{a20} has an appropriate form to implement the NU method.
We obtain the following equations
\ba
\begin{split}
\tilde{\tau}(s)&= 1-s,\\
\sigma(s)&=s(1-s),\\
\tilde{\sigma}(s)&=-\varepsilon^{2}(1-s)^{2} -\alpha^{2}s^2+\beta^2
s(1-s)-s\ell(\ell+1), \label{a21}
\end{split}
\ea
after comparing Eq.\eqref{a20} with Eq.\eqref{a17}.

Furthermore, factorizing
\ba \chi(s)=\phi(s)y(s), \label{a22}
\ea
the equation~\eqref{a17} reduces into the following hypergeometric type equation
\ba \sigma (s)
y^{''}(s)+\tau(s)y^{'}(s)+\bar{\lambda}y(s)=0. \label{a23}
\ea
However, the suitable function $\phi (s)$ need to obey the condition
\ba
\frac {\phi^{'}(s)}{\phi(s)}=\frac {\pi (s)}{\sigma
(s)}, \label{a24}
\ea
with
\ba
\pi(s)=\frac{{\sigma'-\tilde{\tau}}}{2} \pm \sqrt
{\left(\frac{{\sigma'-\tilde{\tau}}}{2}\right)^{2}-\tilde{\sigma}
+k\sigma}, \label{a25}
\ea
and it can be at most first-order polynomial.
As a result, the equation turns to the form of
hypergeometric-type, where $y(s)$ is one of its solutions,
providing that the polynomial
$\bar\sigma(s)=\tilde\sigma(s)+\pi^2(s)+\pi(s)[\tilde\tau(s)-\sigma^{'}(s)]+\pi^{'}(s)\sigma(s)$
can be divided by a factor $\sigma(s)$, i.e., $\bar\sigma/\sigma(s)=\bar \lambda
$. The $\tau (s)$ and $\bar \lambda$ and in
Eq.\eqref{a17} are given by
\ba \bar{\lambda}=k+\pi^{'}(s), \label{a26}
\ea
\ba \tau(s)=\tilde{\tau}(s)+2\pi(s), \label{a262}
\ea
respectively. We obtain $\pi(s)$ in this study as
\ba \pi(s)=\frac{{-s}}{2} \pm \sqrt{s^{2}[a-k]-s[b-k]+c}
\label{a23}
\ea
in which $a=\frac{1}{4}+\varepsilon^{2}+\alpha^{2}+\beta^{2}$, $b=2\varepsilon^{2}+\beta^{2}-\ell(\ell+1)$ and $c=\varepsilon^{2}$.
According to the NU method, the quadratic form under the square-root sign of Eq.\eqref{a23} need to be solved by
setting its discriminant equal to zero. This discriminant provides a
new relation which is solvable for the constant to obtain the roots as
\ba k_{\pm}=(b-2c)
\pm 2\sqrt{c^{2}+c(a-b)}~. \label{a27}
\ea
By substituting the two values of $k_{\pm}$ into Eq.\eqref{a23}, we find the four different expressions of $\pi(s)$ as follows
\ba \pi(s)=\frac{{-s}}{2} \pm \left\{
\begin{array}{l}
(\sqrt{c}-\sqrt {c+a-b})s-\sqrt c ~\text{for}~ k_{+} = (b-2c)+2\sqrt{c^2+c(a-b)} \\
(\sqrt{c}+\sqrt {c+a-b})s-\sqrt c ~\text{for}~ k_{-} = (b-2c)-2\sqrt{c^2+c(a-b)} \\
\end{array} \right.
\label{a25}
\ea
However, one of the above expressions is just appropriate to find the bound state solutions since the $\tau (s)$ must have the negative derivative. Others have no physical meaning. Consequently, the suitable functions for $\pi(s)$
and $\tau(s)$, which satisfy the bound state condition, are
\ba
\pi(s)=\sqrt{c}-s\left[\frac{1}{2}+\sqrt{c}+\sqrt{c+a-b}\right],
\label{a27}
\ea
and
\ba \tau(s)=1+2\sqrt{c}-2s
\left[1+\sqrt{c+a-b}\right], \label{a28}
\ea
for
$
k_{-}=(b-2c)-2\sqrt{c^{2}+c(a-b)}. \label{a29}
$
Moreover, the constant $\bar{\lambda}$ is obtained from Eq.\eqref{a26} as follows:
\ba
\bar{\lambda}=b-2c-2\sqrt{c^2+c(a-b)}-\left[\frac{1}{2}+{\sqrt{c}+\sqrt{c+a-b}}\right].
\label{a30}
\ea
Considering an integer $n_r\geq0$, a unique polynomial solution for the hypergeometric
type equation of degree $n_r$ is obtained if
\ba
\bar{\lambda}=\bar{\lambda}_{n_r}=-n_{r}\tau'-\frac{n_r(n_r-1)}{2}\sigma'',
(n_r=0,1,2...), \label{a31}
\ea
and
$\bar{\lambda}_m\neq\bar{\lambda}_n $ for $m=0,1,2,...,n_{r}-1$, then we have,
\ba
\begin{split}
\bar{\lambda}_{n_{r}}&=b-2c-2\sqrt{c^2+c(a-b)}-\left[\frac{1}{2}+{\sqrt{c}+\sqrt{c+a-b}}\right]\\
 &=2n_r\left[{1 +\left({\sqrt{c}+\sqrt{c+a-b}}
\right)}\right]+n_r(n_r-1). \label{a32}
\end{split}
\ea
This equation can be clearly solved for $c$ with the relation
$c=\varepsilon^2$, hence
\ba
\varepsilon^{2}
=\left[\frac{\beta^{2}-\ell(\ell+1)-1/2-n_r(n_r+1)-(2n_r+1)\sqrt{\frac{1}{4}+\alpha^2+\ell(\ell+1)}}{2n_r+1+2\sqrt{\frac{1}{4}+\alpha^2+\ell(\ell+1)}}
\right]^{2}.~~~\label{a33}
\ea
By using $\varepsilon^{2}$ in Eq.~\eqref{a19} along with Eq.~\eqref{a33}, we get
\ba
M^2-E_{n_{r},\ell}^{2}=\left[\frac{\beta^{2}-\ell(\ell+1)-1/2-n_r(n_r+1)-(2n_{r}+1)\sqrt{\frac{1}{4}+\alpha^2+\ell(\ell+1)}}{n_r+\frac{1}{2}+\sqrt{\frac{1}{4}+\alpha^2+\ell(\ell+1)}}\times\delta\right]^2~~~~
\label{a34}
\ea
The energy eigenvalues $E_{n_r,\ell}$ can be calculated using this result, which is more tricky than the square equation.

Now, we begin to treat the radial eigenfunction. After putting $\sigma(s)$ and $\pi(s)$ into
Eq.\eqref{a24},  we get
\ba \phi (s)=s^{\varepsilon}(1-s)^\kappa, \label{a39}
\ea
where $\kappa=1/2+\sqrt{\frac{1}{4}+\ell(\ell+1)+\alpha^2}$. At the same time $y_n(s)$, the other part of the wave equation, is the hypergeometric-type function and its polynomial solutions are obtained by using
Rodrigues relation for a fixed integer $n$:
\ba y_{n}(s)=\frac{C_{n}}{\rho (s)}
\frac{{d^{n}}}{{ds^{n}}}\left[\sigma^{n}(s)\rho(s) \right],
\label{a40}
\ea
where $C_n$ and $\rho (s)$ denote the constant of normalization and the weight function, respectively. Notice that $\rho (s)$ is also known as the solution of the Pearson
differential equation
\ba (\sigma
\rho )^{'} =\tau \rho. \label{a41}
\ea
The weight function $\rho(s)$ for our problem are obtained as follows
\ba
\rho(s)=(1-s)^{2\kappa-1}s^{2\varepsilon}. \label{a42}
\ea
Substituting the Eq.\eqref{a42} to Eq.\eqref{a40}, we have
\ba
y_{n_{r}}(s)=C_{n_{r}}(1-s)^{1-2\kappa}s^{-2\varepsilon}
\frac{{d^{n_{r}}}}{{ds^{n_{r}}}}\left[{s^{{2\varepsilon}+n_{r}}
(1-s)^{2\kappa-1+n_{r}}}\right]. \label{a43}
\ea
Next, by considering the Jacobi polynomials \cite{Abramowitz}
\ba P_n^{(a,b)}(s)=\frac{(-1)^n}{2^n n!(1-s)^a(1+s)^b}\frac{d^n}{ds^n
}\left[{(1-s)^{a+n}(1+s)^{b+n}}\right], \label{a44}
\ea
we can express
\ba
P_n^{(a,b)}(1-2s)=\frac{1}{n!s^a(1-s)^b}\frac{d^{n}}{ds^{n}}\bigl[s^{a+n}(1-s)^{b+n}\bigr],
\label{a45} \ea
which implies that
\ba \frac{d^n }{ds^n
}\bigl[s^{a+n}(1-s)^{b+n}\bigr]=n!s^{a}(1-s)^{b}
P_n^{(a,b)}(1-2s). \label{a46}
\ea
We can then use this to express $y_{n_{r}}(s)$ on Eq.~\eqref{a40} as
\ba
y_{n_{r}}(s) = C_{n_{r}} P_{n_{r}}^{(2\varepsilon,2\kappa-1)}(s),
\label{a47}
\ea
and then putting Eq.~\eqref{a39} and Eq.~\eqref{a47} into the
Eq.~\eqref{a22} gives
\ba
\chi_{n_{r}}(s)=C_{n_{r}}s^{\varepsilon}(1-s)^\kappa
P_{n_{r}}^{(2\varepsilon,2\kappa-1)}(s). \label{a48}
\ea

From the following expression of Jacobi polynomials \cite{Abramowitz}
\ba
P_n^{(a,b)}(s)=\frac{{\Gamma(n+a+1)}}{{n!\Gamma(a+1)}}
{}_2F_{1} \left({-n,n+a+b+1,1+a;s}\right), \label{a49}
\ea
the equation~\eqref{a48} can be expressed in terms of the hypergeometric functions:
\ba \chi_{n_{r}}
(s)=C_{n_{r}}s^{\varepsilon}(1-s)^{\kappa}\frac{\Gamma (n_{r}+2\varepsilon+1)}{n_{r}!\Gamma (2\varepsilon+1)} {}_2F_{1}
\left({-n_{r},{2\varepsilon}+2\kappa+n_{r},1+{2\varepsilon};s}\right).
\label{a50}
\ea
The normalization constant $C_{n_{r}}$ is determined
via the following condition
\ba \int\limits_0^\infty
|R(r)|^{2}r^{2}dr=\int\limits_0^{\infty} |\chi(r)|^{2}
dr=\frac{1}{2\delta}\int\limits_0^1\frac{1}{s}|\chi (s)|^{2}ds=1.
\label{a51}
\ea
Applying the following integral relation
\cite{Abramowitz}
\ba
\begin{split}
\int\limits_0^1 {(1-z)^{2(\delta+1)}z^{{2\lambda}-1}}&
\biggl[{ {}_2F_{1} (-n_{r},2(\delta+\lambda+1)+n_{r},2\lambda+1;z)} \biggr]^2 dz \\
&= \frac{{n_{r}!(n_{r}+{\delta}+1)\Gamma (n_{r}+{2\delta}+2)\Gamma (2\lambda)\Gamma({2\lambda}+1)}}{{(n_{r}+{\delta}+{\lambda}+1)\Gamma(n_{r}+{2\lambda}+1)\Gamma (2({\delta}+{\lambda}+1)+n_{r})}},
 \label{a52}
\end{split}
\ea
where $\delta > \frac{{-3}}{2}$ and $\lambda>0 $, the normalization constant can be easily obtained as
\ba C_{n_r}=\sqrt{\frac{2\delta n_{r}!(n_{r}+\kappa+\varepsilon)\Gamma(n_r+{2\varepsilon}+2\kappa)\Gamma
({2\varepsilon}+1)}{(n_{r}+\kappa)\Gamma(n_r+{2\varepsilon}+1) \Gamma
(2\varepsilon)\Gamma (n_{r}+2\kappa)}}. \label{a53} \ea

\subsection{Implementation of SUSYQM Method}\label{sr}
In the SUSYQM, the ground state eigenfunction $\chi_{0}(r)$ in Eq.\eqref{a6} is defined by
\ba
\chi_{0}(r)=Nexp\left(-\int W(r)dr\right), \label{a54}
\ea
where $N$ and $W(r)$ are respectively the normalization constant and superpotential function.
The relations between $W(r)$ and $V_{\pm}(r)$, the supersymmetric partner
potentials, are given by \cite {Cooper1,Cooper2}:
\ba V_{\pm}(r)=W^{2}(r)\pm W'(r). \label{a55}
\ea
The Riccati equation~\eqref{a55} have the particular solution
\ba W(r)=\left(F-\frac{Ge^{-2\delta
r}}{1-e^{-2\delta r}}\right), \label{a56}
\ea
in which $F$ and $G$ are arbitrary constants. By employing $V_{-}(r)=V_{\rm eff}(r)-(E^2-M^2)$,
Eqs.~\eqref{a16} and \eqref{a56} are added into the Eq.\eqref{a55}. After that,
by comparing suitable quantities in the right and left hand sides, we get the following relations for $F$ and $G$:
\ba F^{2}=4
\delta^{2}\varepsilon^{2}, \label{a57}
\ea
\ba 2FG+2\delta
G=4\delta^2\beta^2- 4\delta^{2}\ell(\ell+1), \label{a58}
\ea
\ba
G^{2}-2\delta G=4\delta^{2}\alpha^{2}+4\delta^{2}\ell(\ell+1). \label{a59}
\ea
Applying an extreme condition for the wave functions, we find that $F<0$ and $G>0$. Then, solving Eq.\eqref{a59} leads to
\ba
G=\frac{2\delta\pm\sqrt{4\delta^{2}+16\delta^{2}(\alpha^{2}+\ell(\ell+1))}}{2}=\delta\pm
2\delta\sqrt{\frac{1}{4}+\ell(\ell+1)+\alpha^{2}}. \label{a60}
\ea
When consider $G>0$ , from Eqs.\eqref{a58} and \eqref{a59} we find
\ba
2FG+G^{2}=4\delta^{2}(\alpha^{2}+\beta^{2}), \label{a61} \ea
or
\ba
F=-\frac{G}{2}+\frac{2\delta^{2}(\alpha^{2}+\beta^{2})}{G}.
\label{a62}
\ea
Then, from Eqs.~\eqref{a57} and~\eqref{a62}, we obtain
\ba
\varepsilon^{2}=\frac{1}{4\delta^{2}}\left[-\frac{\delta+2\delta\sqrt{\frac{1}{4}+\ell(\ell+1)+\alpha^{2}}}{2}-\frac{2\delta^2{(\alpha^{2}+\beta^{2})}}{\delta+2\delta\sqrt{\frac{1}{4}+\ell(\ell+1)+\alpha^{2}}}\right]^2.\label{a63}
\ea
By putting the Eq.~\eqref{a63} into Eq.~\eqref{a19}, for the energy spectrum, we find,
\ba
M^{2}-E^{2}=\left[-\frac{\delta+2\delta\sqrt{\frac{1}{4}+\ell(\ell+1)+\alpha^{2}}}{2}-\frac{2\delta{(\alpha^{2}+\beta^{2})}}{1+2\sqrt{\frac{1}{4}+\ell(\ell+1)+\alpha^{2}}}\right]^2.\label{a64}
\ea

In the limit $r\rightarrow\infty$, $W(r)\rightarrow$ $F$. Substituting the Eq.\eqref{a56} into Eq.\eqref{a55} allows us to represent
\ba V_{+}(r)&=&
W^{2}(r)+W'(r)=\left[F^{2}-\frac{(2FG-2\delta G)e^{-2\delta
r}}{1-e^{-2\delta r}}+\frac{(G^2+2\delta G)e^{-4\delta
r}}{(1-e^{-2\delta r})^2}\right], \label{a66}
\ea
\ba
V_{-}(r)=W^{2}(r)-W'(r)=\biggl[F^2-\frac{(2FG+2\delta G)e^{-2\delta
r}}{1-e^{-2\delta r}}+\frac{(G^2-2\delta G)e^{-4\delta
r}}{(1-e^{-2\delta r})^2}\biggr]. \label{a65}
\ea
The potentials $V_{\pm}(r)$ are different with each other by additive constant. However, these have functional form similarity, which are named as the invariant
potentials \cite{Gendenshtein1, Gendenshtein2}. Their invariant forms are
\ba
\begin{split}
R(G_1) &= V_{+}(G,r)-V_{-}(G_1,r) \\
 &=\left[F^{2}-F_{1}^{2}\right] \\
&=\left[-\frac{G}{2}+\frac{2\delta^{2}(\alpha^{2}+\beta^{2})}{G}\right]^{2}+\left[\frac{G+2\delta}{2}+\frac{2\delta^{2}(\alpha^{2}+\beta^{2})}{G+2\delta}\right]^2,
\label{a68}
\end{split}
\ea
\ba
\begin{split}
R(G_{i})&=V_{+}[G+2\delta \times (i-1),r]-V_{-}[G+2\delta \times (i) ,r] \\
&=\left(-\frac{G+2\delta \times (i-1)}{2}+\frac{2(\alpha^{2}+\beta^{2})\delta^2}{G+2\delta \times (i-1)}\right)^{2}
-\left(\frac{G+2\delta \times i}{2}-\frac{2(\alpha^{2}+\beta^{2})\delta^{2}}{G+2\delta \times i}\right)^2,~~~~~~~ \label{a69}
\end{split}
\ea
where $R(G_i)$ does not depend on $r$. We continue by using $\,G_{n_r}=G_{n_r-1} +2\delta =G+2n_r\delta$, and then obtain all the discrete spectrum of Hamiltonian $\,H_{-}(G)$ as follows
\ba
E^{2}_{n_r}=E^2_{0}+\sum_{i=1}^{n_r}R(G_i)\label{a70}
\ea
\ba
\begin{split}
E^{2}_{n_r}&=M^2-\left(-\frac{G}{2}+\frac{2\delta^{2}(\alpha^{2}+\beta^{2})}{G}\right)^{2}+\left(-\frac{G}{2}+\frac{2\delta^{2}(\alpha^{2}+\beta^{2})}{G}\right)^{2}\\
&-\left(-\frac{G+2\delta}{2}+\frac{2\delta^{2}(\alpha^{2}+\beta^{2})}{G+2\delta}\right)^{2}+ \left(-\frac{G+2\delta}{2}+\frac{2\delta^{2}(\alpha^{2}+\beta^{2})}{G+2\delta}\right)^{2}\\
& - \cdots -\\
&-\biggl(-\frac{G+(n_r-1)2\delta}{2}+\frac{2\delta^{2}(\alpha^{2}+\beta^{2})}{G+2(n_r-1)\delta}\biggr)^{2}+\left(-\frac{G+2(n_r-1)\delta}{2}+\frac{2\delta^{2}(\alpha^{2}+\beta^{2})}{G+2(n_r-1)\delta}\right)^2\\
&-\left(-\frac{G+2
n_r\delta}{2}+\frac{2\delta^{2}(\alpha^{2}+\beta^{2})}{(G+2
n_r\delta)}\right)^{2}.
\end{split} \label{a71}
\ea
As a result, we obtain
\ba E^{2}_{n_{r}l}
&=&M^2-\left(-\frac{G+2\cdot
n_r\delta}{2}+\frac{2\delta^{2}(\alpha^{2}+\beta^{2})}{G+2
n_r\delta} \right)^2. \label{a72}
\ea
Finally, putting $G$ in Eq.~\eqref{a60} into the
Eq.~\eqref{a72}, for energy spectrum, we get the following
form:
\ba
\begin{split}
M^{2}-&E_{n_{r},\ell}^{2}=\delta^2\left[-\sqrt{\frac{1}{4}+\alpha^{2}+\ell(\ell+1)}- n-\frac{1}{2}  +\frac{\left(\alpha ^2+\beta ^2\right)}{\sqrt{\frac{1}{4}+\alpha^{2}+\ell(\ell+1)}+n+\frac{1}{2}}\right]^2\\
&=\delta^2\left[\frac{\beta^{2}-\ell(\ell+1)-1/2-n_r(n_r+1)-(2n_{r}+1)\sqrt{\frac{1}{4}+\alpha^{2}+\ell(\ell+1)}}{n_r+\frac{1}{2}+\sqrt{\frac{1}{4}+\alpha^{2}+\ell(\ell+1)}}\right]^2. \end{split}
\label{a73}
\ea
This form is exactly identical with the one we get by the NU method in ~\eqref{a34}.

Furthermore, by inserting the superpotential~\eqref{a56} into the Eq.~\eqref{a54}, the radial eigenfunction $\chi_{0}(r)$
 is calculated as
\ba
\begin{split}
\chi_{0}(r)&=N \exp\left[-\int W(r)dr\right]=N \exp\left[\int\left(-F+\frac{Ge^{-2\delta r}}{1-e^{-2\delta r}}\right)dr\right] \\
&=Ne^{Fr} \exp\left[\frac{G}{\delta}\int\frac{d(1-e^{-2\delta
r})}{1-e^{-2\delta r}}\right]\\
&= N e^{Fr}(1-e^{-2\delta
r})^{\frac{G}{2\delta}}. \label{a67}
\end{split}
\ea
When $r\rightarrow 0$, we can see that $\chi_{0}(r)\rightarrow0$ and $G>0$. When $r\rightarrow \infty$,
$\chi_{0}(r)\rightarrow0$ and $F<0$.

\section{Particular cases}\label{pc}
After examining the bound state solutions of any $l$-state KG equation with a Class of Yukawa plus Manning-Rosen potentials, now we discuss some particular cases below. By adjusting potential parameters for each cases, some familiar potentials, which are useful for other physical systems, can be obtained.
\begin{enumerate}
  \item Setting $V_{0}$ and $V'_{0}$ to zero, the potential turns to central Manning-Rosen potential. In this case, the energy spectrum equation is
\ba
M^{2}-E^{2}=\delta^2\left[\frac{\gamma^{2}-\ell(\ell+1)-1/2-n_r(n_r+1)-(2n_{r}+1)\sqrt{\frac{1}{4}+\alpha^{2}+\ell(\ell+1)}}{n_r+\frac{1}{2}
+\sqrt{\frac{1}{4}+\alpha^{2}+\ell(\ell+1)}}\right]^2 \label{a74}
\ea
where
\ba
\gamma=\frac{\sqrt{2(E+M)V_{02}}}{2\delta}=\sqrt{\frac{A(E+M)}{M}}. \label{a75}
\ea
This result is the same with the expression for bound state obtained in Ref.~\cite{Dong3} (see, Eq.(18) of Ref.~\cite{Dong3}).
The corresponding wave function is
\ba
\chi_{n_{r},\ell}(s)=N_{n_{r},l}s^{\frac{\sqrt{M^2-E^2}}{2\delta}}(1-s)^{\frac{1}{2}+\sqrt{\frac{1}{4}+\ell(\ell+1)+\alpha^2}}
P_{n_{r}}^{\bigl(\frac{\sqrt{M^2-E^2}}{\delta},2\sqrt{\frac{1}{4}+\ell(\ell+1)+\alpha^2}\bigr)}(s).
\ea

\item Setting $V_{01}$ and $V_{02}$ to zero, i.e., $\eta=1$ and $A=0$, the potential turns to a class of Yukawa potential given in Eq.\eqref{potCY}. For this potential, the energy spectrum equation is obtained as
\ba
M^{2}-E^{2}=\delta^2\left[\frac{\xi^{2}-\ell(\ell+1)-1/2-n_r(n_r+1)-(2n_{r}+1)\sqrt{\frac{1}{4}+\zeta^{2}+\ell(\ell+1)}}{n_r+\frac{1}{2}
+\sqrt{\frac{1}{4}+\zeta^{2}+\ell(\ell+1)}}\right]^2 \label{aclasYuk}
\ea
where
\ba
\xi=\frac{\sqrt{4\delta V_{0}(E+M)}}{2\delta}, \label{aclasYukpar1}
\zeta=\sqrt{-2 V'_{0}(E+M)}. \label{aclasYukpar2}
\ea
The corresponding wave function is
\ba
\chi_{n_{r},\ell}(s)=N_{n_{r},l}s^{\frac{\sqrt{M^2-E^2}}{2\delta}}(1-s)^{\frac{1}{2}+\sqrt{\frac{1}{4}+\ell(\ell+1)+\zeta^2}}
P_{n_{r}}^{\bigl(\frac{\sqrt{M^2-E^2}}{\delta},2\sqrt{\frac{1}{4}+\ell(\ell+1)+\zeta^2}\bigr)}(s).
\ea

  \item Setting $\eta=1$ and $V_{0}=V'_{0}=0$, the potential turns to Hulten potential. In this case, the energy spectrum equation is
\ba
M^{2}-E^{2}=\delta^2\left[\frac{\gamma^{2}-\ell(\ell+1)-1/2-n_r(n_r+1)-(2n_{r}+1)\sqrt{\frac{1}{4}+\ell(\ell+1)}}{n_r+\frac{1}{2}
+\sqrt{\frac{1}{4}+\ell(\ell+1)}}\right]^2~~~~ \label{a744}
\ea
This result is the same with the expression obtained in Eq.(50) of Ref.~\cite{Ahmadov3} under the choice of $S(r)=V(r)$.
The corresponding wave function is
\ba
\chi_{n_{r},\ell}(s)=N_{n_{r},l}s^{\frac{\sqrt{M^2-E^2}}{2\delta}}(1-s)^{\frac{1}{2}+\sqrt{\frac{1}{4}+\ell(\ell+1)}}
P_{n_{r}}^{\bigl(\frac{\sqrt{M^2-E^2}}{\delta},2\sqrt{\frac{1}{4}+\ell(\ell+1)}\bigr)}(s).
\ea
  \item If the parameters $V_{01}$, $V_{02}$ and $V_{04}$ are set to zero, i.e., $\eta=1,A=V'_{0}=0$, then we have the central Yukawa potential. The energy spectrum equation for this case is
\ba
M^{2}-E^{2}=\delta^2\left[\frac{\xi^{2}-\ell(\ell+1)-1/2-n_r(n_r+1)-(2n_{r}+1)\sqrt{\frac{1}{4}+\ell(\ell+1)}}{n_r+\frac{1}{2}+\sqrt{\frac{1}{4}+\ell(\ell+1)}}\right]^2 \label{a76}
\ea
where $\xi$ is given in Eq.~\eqref{aclasYukpar1}.
This result is the same with the expression for the constant mass case obtained in Ref.~\cite{Wang2015}. One can easily see this by setting $q=1$ and $\alpha\rightarrow\delta$ in Eq.(39) of Ref.~\cite{Wang2015}.
The corresponding wave function is
\ba
\chi_{n_{r},\ell}(s)=N_{n_{r},\ell}s^{\frac{\sqrt{M^2-E^2}}{2\delta}}(1-s)^{\frac{1}{2}+\sqrt{\frac{1}{4}+\ell(\ell+1)}}
P_{n_{r}}^{\bigl(\frac{\sqrt{M^2-E^2}}{\delta},2\sqrt{\frac{1}{4}+\ell(\ell+1)}\bigr)}(s).
\ea

  \item If the parameters $V_{01}$, $V_{02}$ and $V_{03}$ are set to zero, i.e., $\eta=1,A=V_{0}=0$, then we have the inversely quadratic Yukawa potential. The energy spectrum equation for this case is
\ba
M^{2}-E^{2}=\delta^2\left[\frac{-\ell(\ell+1)-1/2-n_r(n_r+1)-(2n_{r}+1)\sqrt{\frac{1}{4}+\zeta^{2}+\ell(\ell+1)}}{n_r+\frac{1}{2}+\sqrt{\frac{1}{4}+\zeta^{2}+\ell(\ell+1)}}\right]^2 \label{a762}
\ea
where $\zeta$ is given in Eq.~\eqref{aclasYukpar2}.
The corresponding wave function is given by
\ba
\chi_{n_{r},\ell}(s)=N_{n_{r},\ell}s^{\frac{\sqrt{M^2-E^2}}{2\delta}}(1-s)^{\frac{1}{2}+\sqrt{\frac{1}{4}+\zeta^{2}+\ell(\ell+1)}}
P_{n_{r}}^{\bigl(\frac{\sqrt{M^2-E^2}}{\delta},2\sqrt{\frac{1}{4}+\zeta^{2}+\ell(\ell+1)}\bigr)}(s).
\ea
  \item If $\delta\rightarrow 0$ in Eq.\eqref{a76}, the potential reduces to  Coulomb-like potential, $V(r)=-V_{0}/r$, and the corresponding energy spectrum is obtained as
\ba
M^{2}-E^{2}=\left[\frac{ V_0 (E+M)}{(\ell+n+1)}\right]^2 \Rightarrow E=M\frac{(n+\ell+1)^2- V_0^2 }{(n+\ell+1)^2+ V_0^2}\label{a78}
\ea
and this result is the same with Eq.(51) of Ref.~\cite{Wang2015} and consistent with those
results in Ref.~\cite{Ikhdair09}.
  \item For $\ell=0$ (the s-wave case), the centrifugal term in Eq.\eqref{a15} disappears because ${\ell(\ell+1)\delta^2\frac{e^{-2\delta r}}{(1-e^{-2\delta r})^2}}=0$ and the equation turns to the $s$-wave KG equation. We can obtain its corresponding energy spectrum and radial wave functions from Eq.\eqref{a34} and Eq.\eqref{a50} by setting $l=0$, respectively. The energy spectrum equation is given by
      \ba
M^2-E^{2}=\delta^2\left[\frac{\beta^{2}-1/2-n_r(n_r+1)-(2n_{r}+1)\sqrt{\frac{1}{4}+\alpha^2}}{n_r+\frac{1}{2}+\sqrt{\frac{1}{4}+\alpha^2}}\right]^2.~~~~~
\label{a79}
\ea

\end{enumerate}

\section{Numerical Evaluation}\label{nr}
\begin{figure}[b]
    \begin{center}
\includegraphics[scale=0.40]{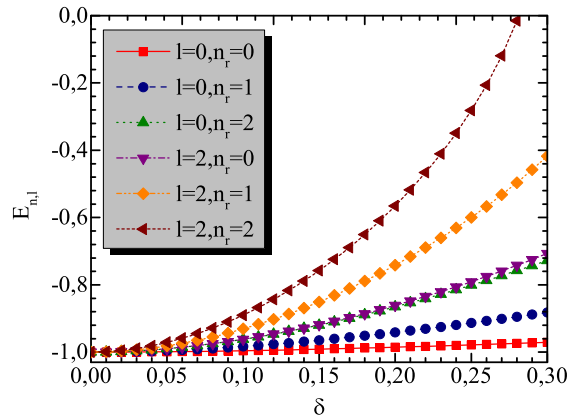}
     \end{center}
          \vspace{-4mm}
\caption{ The variation of $E_{n_r,l}$ with respect to $\delta$ for given $n_r$ and $\ell =0,2$.}
\label{fig:Edelta}
\end{figure}

\begin{figure}[h]
    \begin{center}
\includegraphics[scale=0.40]{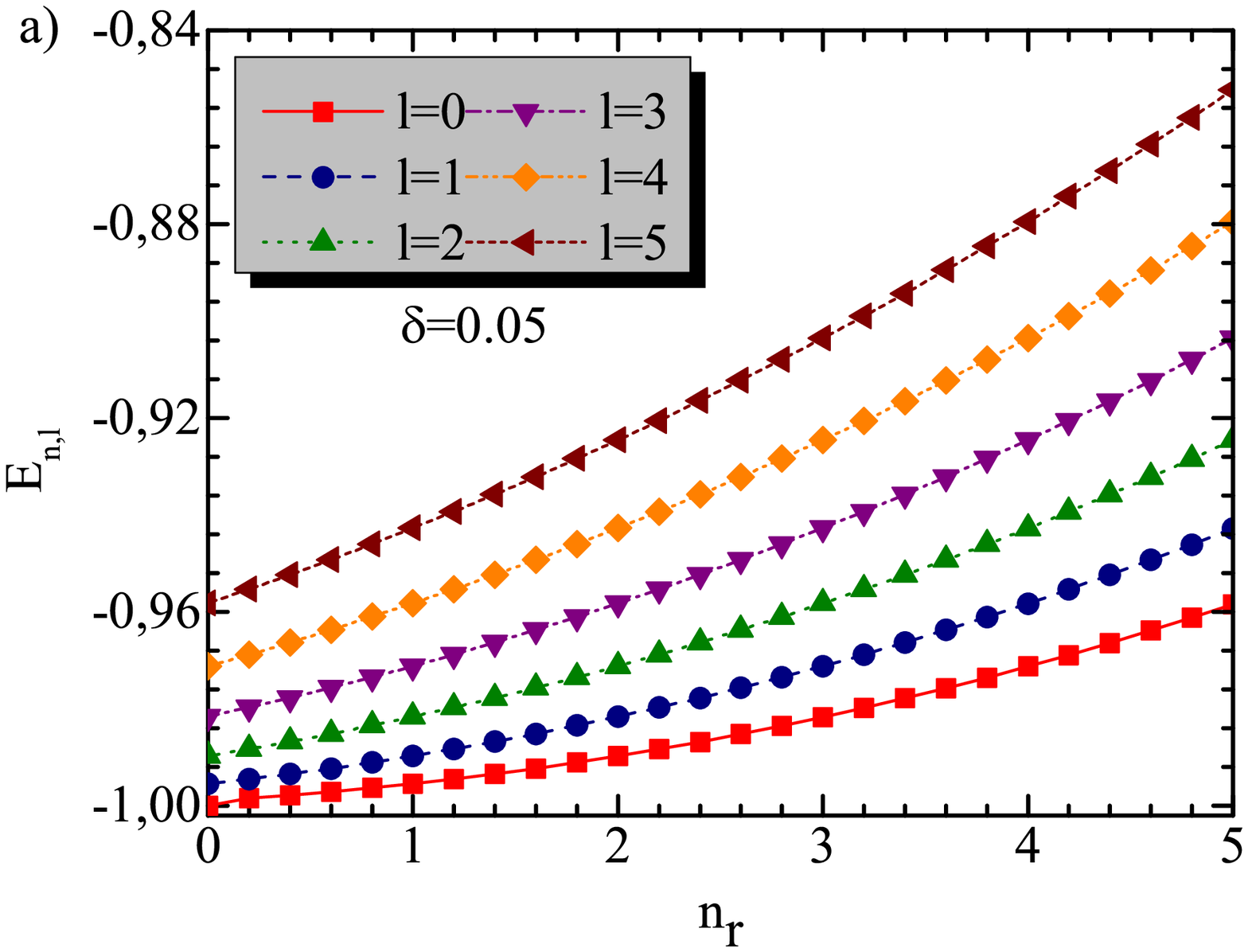}
\includegraphics[scale=0.40]{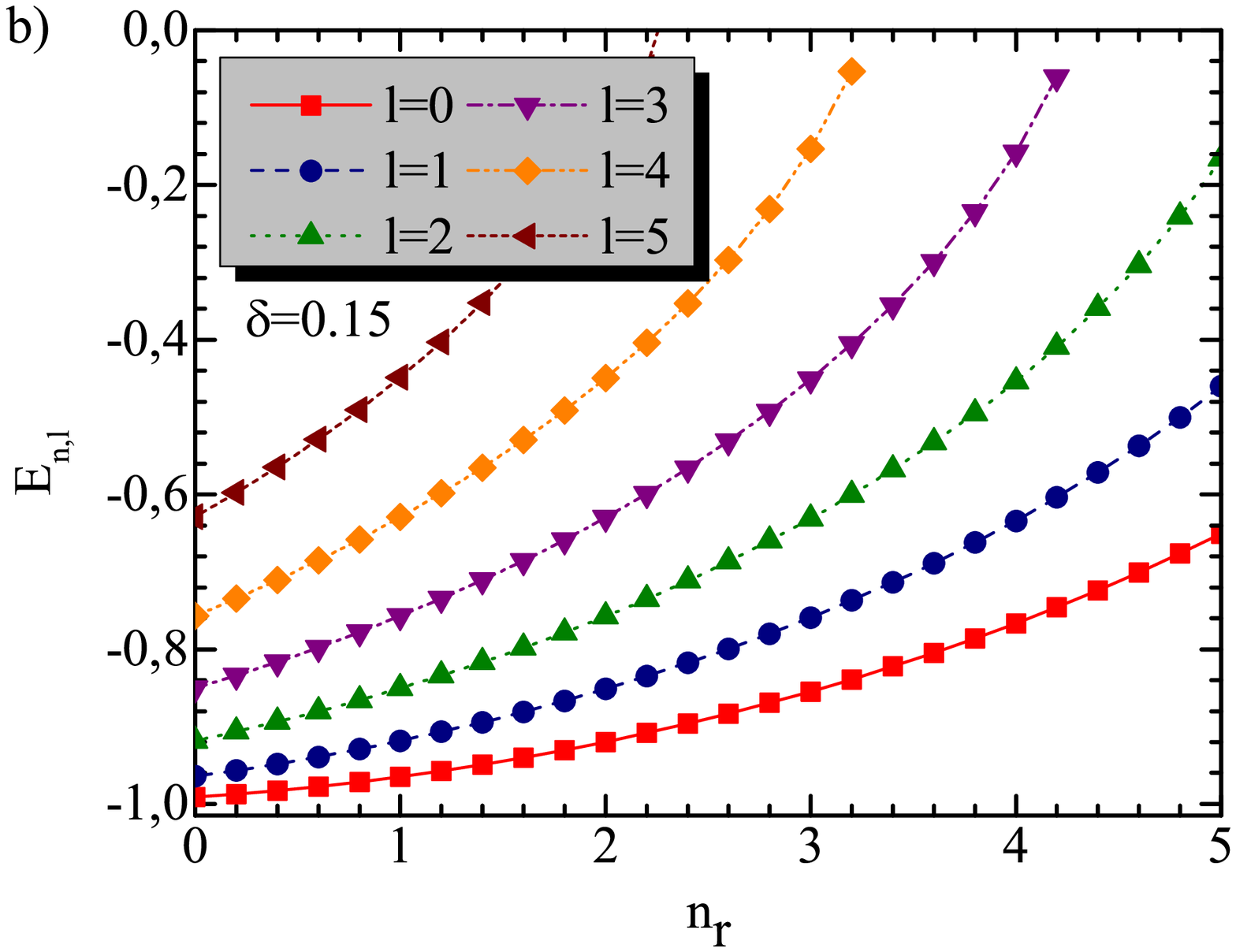}
     \end{center}
         \vspace{-4mm}
\caption{The variation of $E_{n_r,\ell}$ with respect to $n_r$ for given $l$ and a) $\delta=0.05$  and b) $\delta=0.15$.}
\label{fig:Enr}
\end{figure}

In this section, we present the numerical evaluation for the bound state solutions of the $\ell$-wave KG equation with the Manning-Rosen plus a Class of Yukawa potentials. We analyze the dependency of energy levels E on the potential parameter $\delta$ and quantum number $n_r$ for given $l$ as shown in Fig.~\ref{fig:Edelta} and~\ref{fig:Enr}. During our numeric calculation, for simplicity some of the parameters are fixed as follows: $V_0=1$, $V'_0=0.1$,  $M=1$,  $\eta=0.75$ and $A=2b$. We also use the natural units here ($\hbar=c=1$).
We plot the energy levels E in Fig.~\ref{fig:Edelta} as respect to the potential parameter $\delta$ in the range from 0 to 0.30 for $l=0,2$ and $n_r=0,1,2$.
It is seen that the energy levels $E$ have very little variation for an interval of $\delta\in[0,0.05]$ and then continue to rapidly increase with increments of $\delta$. In Fig.~\ref{fig:Enr}(a) and (b), the energy levels $E$ ($\ell=0,..,5$) are plotted as a function of the quantum number $n_r$ in the range from 0 to 5 for $\delta=0.05$ and $0.15$, respectively. For any $\ell$, the energy level E increases with the increment of $n_r$.

\begin{figure}[h]
    \begin{center}
\includegraphics[scale=0.40]{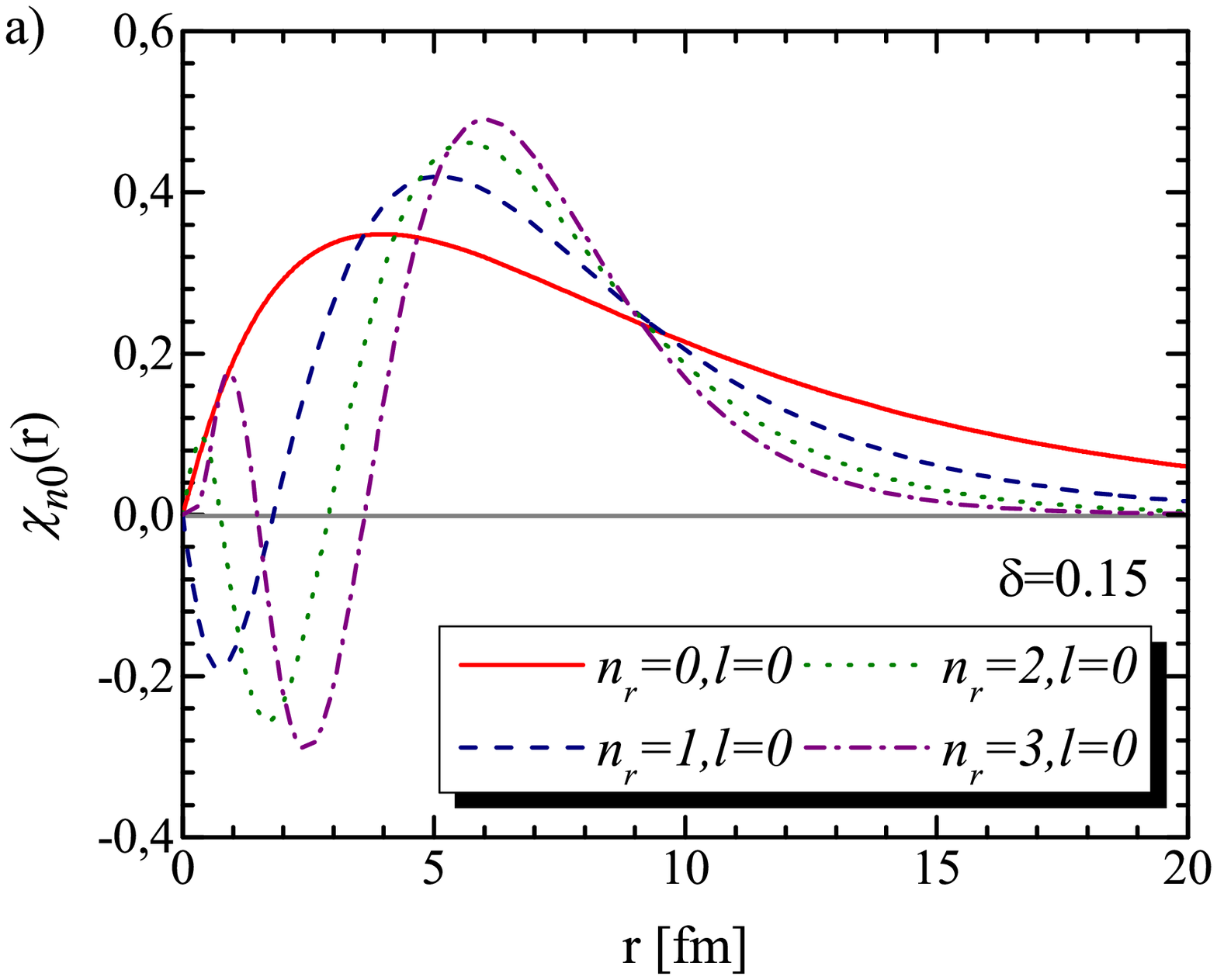}
\includegraphics[scale=0.40]{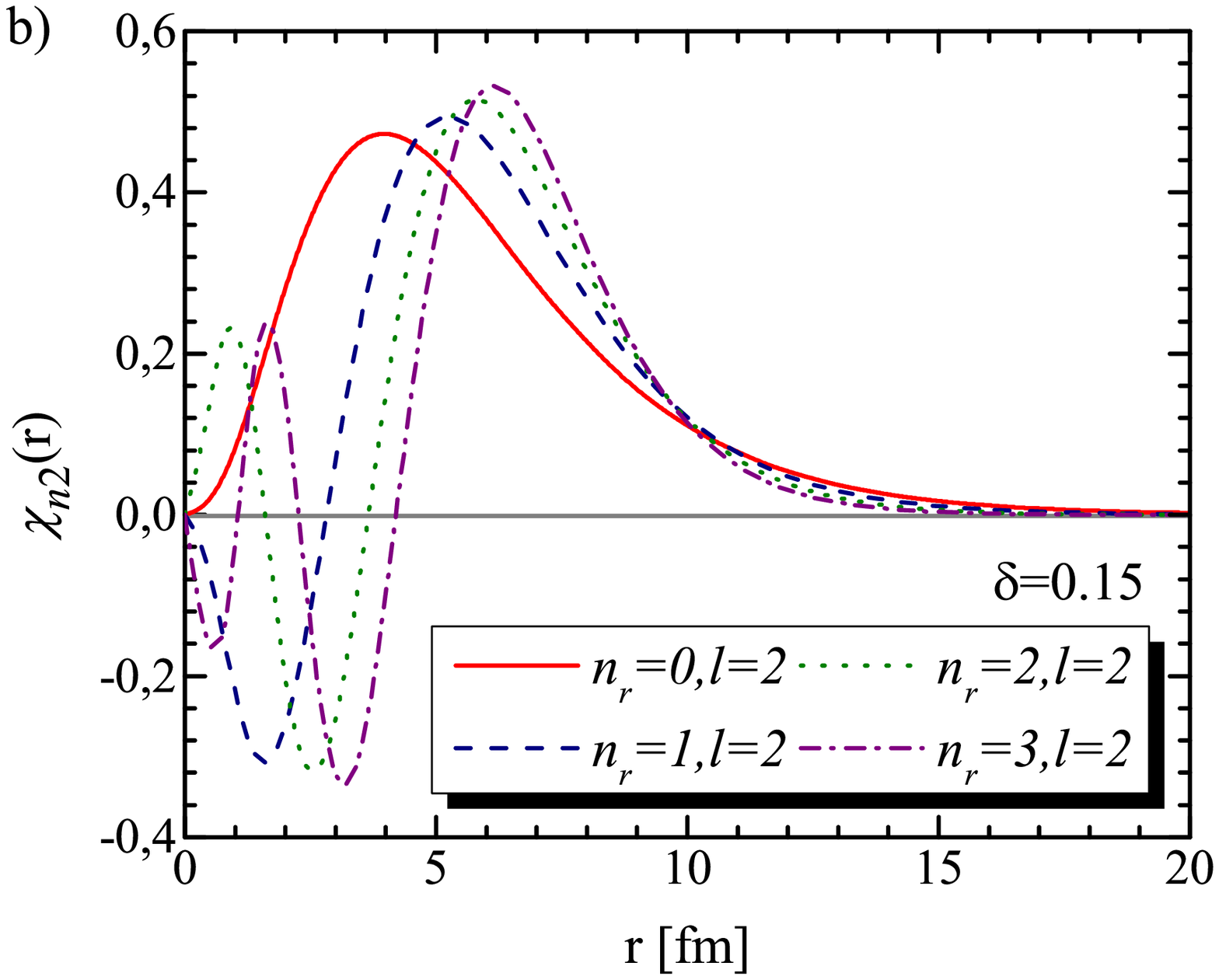}
     \end{center}
     \vspace{-4mm}
\caption{The variation of the normalized wave function $\chi_{n_r,\ell}(r)$ with respect to $r$ for a) $\ell=0$ and b) $\ell=2$.}
\label{fig:wave}
\end{figure}

Figure~\ref{fig:wave} shows total radial wave functions $\chi_{n_r,l}(r)$ as a function of position $r$ varied in the range from 0 to 20 fm for different quantum states of $n = 0, 1, 2, 3$ and $\ell = 0, 2$.
As a mere illustration, the parameter $\delta$ is kept fixed, $\delta=0.15$. Clearly from the figure, the wave functions have $n$ nodes. The position dependence of the potential strength, i.e., $V_0$, $V'_0$, $\eta$, $A$, does not change the number of radial nodes but it affects the wavelength and magnitude of the corresponding wave functions.
\begin{table}[h]
\caption{Bound state energy levels of $1s, 2s, 2p, 3p, 3d, 4p, 4d$ and $4f$ states for given values of $\delta$. The principal quantum number is defined by $n =n_r+\ell+1$.}\label{table:BS}
\centering
\begin{tabular}{ccccccccc}
\hline
$\delta$ & 1s & 2s & 2p& 3p& 3d& 4p& 4d & 4f\\
\hline
0.05& -0.995440 & -0.989722 & -0.981633 & -0.971171 & -0.958218 & -0.958249& -0.942728& -0.924535\\
0.10& -0.983156 & -0.961884 & -0.930941 & -0.890279 & -0.837997 & -0.838488& -0.772955& -0.691402\\
0.15& -0.964688 & -0.919695 & -0.851356 & -0.759000 & -0.631085 & -0.633957& -0.453749& -0.158160\\
0.20& -0.941123 & -0.865398 & -0.743352 & -0.570175 & -0.381591 & -0.300651&  & \\
\hline
\end{tabular}
\end{table}

In Table~\ref{table:BS}, we list bound state energy levels of states $1s, 2s, 2p, 3p, 3d, 4p, 4d, 4f$ for various values of $\delta$.

\section{Concluding remarks}\label{cr}
In this paper, we have employed two alternative methods, the NU and SUSYQM methods, to obtain the bound state solutions of the KG equation in case of equal scalar and vector of Manning-Rosen plus a class of Yukawa potentials by applying the improved approximation scheme to deal with the centrifugal term. One of the main objectives of the current work is to check the validity of the obtained results, as well as to check the legitimacy and the general principles of SUSYQM. We have obtained analytical expression of energy eigenvalues and normalized wave function of a mentioned quantum system for any quantum states $\ell$ and $n_r$. The same expressions are obtained by both methods. It is clear that the bound state solutions are more stable for a class of Yukawa plus Manning-Rosen potentials than the separated cases. The energy spectrum is sensitive with regards to the potential parameter $\delta$ as well as quantum numbers $n_r$ and $\ell$. The wave functions have $n$ nodes. The position dependence of the strength of potential does not change the number of radial nodes but it affects the wavelength and magnitude of the corresponding wave functions.

We have also discussed some special cases, central Yukawa potential, inversely quadratic Yukawa potential, Manning-Rosen potential, Hulten potential, Coulomb-like potential and $s$-wave solution, obtained by adjusting some potential parameters. We have shown that these are consistent with those results in previous works.

The methods used in this work are the systematic ones, and in many cases, they are two of the most reliable techniques in this area. In particular, the potential which is the linear combination of Manning-Rosen and Yukawa potentials can be one of the important exponential potentials and it deserves special concern in many branches of physics, especially in the atomic, condensed matter, particle and nuclear physics.

\appendix
\section{SUSYQM Method}
For $N=2$ in SUSYQM, it is possible to define two nilpotent operators, $Q$ and $Q^{\dagger}$. They satisfy the following
anti-commutation relations:
\ba
\begin{split}
&\{ Q\, ,\, Q\}=0,\, \{Q^{\dagger},Q^{\dagger} \} =0, \\
& \{ Q,\, Q^{\dagger} \}=H.
\end{split}
\ea
Here $H$ is the supersymmetric Hamiltonian operator and conventionally $Q=\left(\begin{array}{cc}{0}&{0} \\
{A^{-}} & {0} \end{array}\right)$ and $Q^{\dagger}
=\left(\begin{array}{cc} {0} & {A^{+}} \\ {0} & {0}
\end{array}\right)$. The $Q$ and $Q^{\dagger} $ are also known as the supercharges operators. Here $A^{-} $ is
bosonic operator and $A^{+}$ is its adjoint. In terms of these operators, the Hamiltonian $H$ can be defined as
\cite{Cooper1,Cooper2}:
\ba H=\left(\begin{array}{cc}{A^{+}
A^{-}}&{0} \\ {0}&{A^{-}A^{+}} \end{array}\right)\,
=\left(\begin{array}{cc} {H_{-} } & {0} \\ {0} & {H_{+} }
\end{array}\right),
\ea
where the $H_{\pm}$ are named as the Hamiltonian of supersymmetric-partner. Note also that $Q$ and
$Q^{\dagger}$ operators commute with $H$. If we have zero ground state energy for $H$ (i.e.
$E_{0}=0$), we can always represent the Hamiltonian as a product of a linear differential
operators pairs in a factorable form. Therefore, the ground state $\psi _{0} (x)$  obeys the Schr\"{o}dinger equation as follows:
\ba H\psi_{o}(x)=-\frac{\hbar^{2}
}{2m} \frac{d^{2} \psi_{0}}{dx^{2}}+V(x)\psi_{0} (x)=0, \ea
hence
\ba V(x)=\frac{\hbar^{2}}{2m} \frac{\psi ''_{0} (x)}{\psi _{0}(x)}.
\ea
This result makes us possible to globally reconstruct the above potential from the information of its ground state wave function that contain zero nodes. Hence, factorizing of $H$ is quite easy by using the following ansatz \cite{Cooper1,Cooper2}:
\ba H_{-}=-\frac{\hbar^{2}}{2m} \frac{d^{2}}{dx^{2}}+V(x)=A^{+}A^{-}
\ea
where
\ba A^{-} =\frac{\hbar}{\sqrt{2m}} \frac{d}{dx}+W(x)\,, \,
A^{+}=-\frac{\hbar}{\sqrt{2m}} \frac{d}{dx}+W(x).
\ea
After that, the Riccati equation for
$W(x)$ can be written as
\ba
V_{-}(x)=W^{2}(x)-\frac{\hbar}{\sqrt{2m}} W'(x).
\ea
Solving for $W(x)$ from this equation, we can express it in terms of $\psi_{0}(x)$ by
\ba W(x)=-\frac{\hbar}{\sqrt{2m}}
\frac{\psi'_{0}(x)}{\psi_{0}(x)}.
\ea
We obtain this solution by noticing that when $A^{-}\psi _{0} (x)=0$ is satisfied, we
have $H\psi_{0}=A^{+} A^{-} \psi_{0}=0
\,.$ We then introduce the operator $H_{+}
=A^{-} A^{+} $ which is written by reversing the order of the $H^{-}$ components.
After a bit simplification, we find that $H_{+}$ is nothing but the
Hamiltonian for new potential $V_{+}(x)$.
\ba H_{+} =-\frac{\hbar^{2}}{2m} \frac{d^{2}}{dx^{2}}+V_{+} (x)\, \,
\, ,\, \, \, \, V_{+}(x)=W^{2}(x)+\frac{\hbar}{\sqrt{2m}} W'(x).
\ea
We call $V_{\pm}(x)$ as supersymmetric
partner potentials. For example, when the ground state energy of $H_{1}$ is $E^{1}_{0}$ with eigenfunction $\psi
_{0}^{1}$, from Eq.(A.5) we can always write
\ba H_{1}=-\frac{\hbar^{2}}{2m}
\frac{d^{2}}{dx^{2}}+V_{1}(x)=A^{+} A^{-}+E_{0}^{1}, \ea
where
\ba
\begin{array}{l} {A_{1}^{-}=\frac{\hbar}{\sqrt{2m}}
\frac{d}{dx}+W_{1}(x)\, ,\, \, A_{1}^{+}=-\frac{\hbar}{\sqrt{2m}}
\frac{d}{dx}+W_{1}(x),}
\\ V_{1}(x)=W_{1}^{2}(x)-\frac{\hbar}{\sqrt{2m}} W'_{1}(x)+E_{0}^{1}, W_{1}(x)=-\frac{\hbar}{\sqrt{2m}}\frac {d \ln \psi_{0}^{1}}{dx} \,.\end{array}
\ea
The SUSY partner Hamiltonian is defined by \cite{Cooper1,Cooper2}
\ba
H_{2} =A_{1}^{-} A_{1}^{+} +E_{0}^{1}=-\frac{\hbar ^{2} }{2m}
\frac{d^{2} }{dx^{2} } +V_{2} (x),
\ea
where
\ba
\begin{split}
V_{2}(x)&=W_{1}^{2}(x)+\frac{\hbar}{\sqrt{2m}}W'_{1}(x)+E_{0}^{1} \\
&= V_{1}(x)+\frac{2\hbar}{\sqrt{2m}}
W'_{1}(x)=V_{1}(x)-\frac{\hbar^2}{m} \frac{d^{2}}{dx^{2}} (\ln
\psi_{0}^{(1)}).
\end{split}
\ea
Using Eq.(A.12), for $H_{1}$ and $H_{2}$, the energy eigenvalues and
eigenfunctions are obtained as
\ba
E_{n}^{2} =E_{n+1}^{1} \, ,\, \, \, \, \psi_{n}^{2}=[E_{n+1}^{1}-E_{0}^{1}]^{-\frac{1}{2}}A_{^{1}}^{-} \psi_{n+1}^{1}\, ,\, \, \, \,
\psi_{n+1}^{1}=[E_{n}^{2} -E_{0}^{1}
]^{-\frac{1}{2}}A_{^{1}}^{+}\psi_{n}^{2}.
\ea
Here $E_{n}^{m}$ represents the energy eigenvalue, where $n$ and $m$ denote the energy level and the  $m$'th Hamiltonian $H_{m}$, respectively. Hence, it is clear that if $H_{1}$ has $p\ge 1$ bound states with
corresponding eigenvalues $E_{n}^{1}$, as well as eigenfunctions $\psi_{n}^{1}$ defined in
$0<n<p$, then we can always generate a hierarchy of $(p-1)$
Hamiltonians, i.e., $H_{2}, H_{3} \, ,\, ...,\, H_{p}$ such that
the $(H_{m})$ has the
same spectrum of eigenvalue as $H_{1}$, apart from the fact that the first $(m-1)$
eigenvalues of $H$ are absent in $H$~\cite{Cooper1,Cooper2}:
\ba
H_{m} =A_{m}^{+} A_{m}^{-} +{\rm \; E}_{{\rm m-1}}^{{\rm
1}}=-\frac{\hbar^{2}}{2m} \frac{d^{2} }{dx^{2} } +V_{m} (x),
\ea
where
\ba A_{m}^{-}=\frac{\hbar}{\sqrt{2m}} \frac{d}{dx}+W_{m}(x)\,
,\, \, \, \, W_{m}(x)=-\frac{\hbar}{\sqrt{2m}} \frac{d\ln
\psi_{0}^{(m)}}{dx},  (m=2\,\,3\,\, 4,\,\, \cdots \, \, p).
\ea
We also have
\ba \begin{array}{l} E_{n}^{(m)}=E_{n+1}^{(m-1)}=\cdots =E_{n+m-1}^{1} \,, \\ \psi _{n}^{(m)} =[E_{n+m-1}^{1}-E_{m-2}^{1}]^{-\frac{1}{2}} \cdots [E_{n+m-1}^{1}-E_{0}^{1}]^{-\frac{1}{2}}A_{m-1}^{-} \cdots A_{1}^{-} \psi^{1}_{n+m-1}, \,\,\,\, \\
V_{m}(x)=V_{1}(x)-\frac{\hbar^2}{m} \frac{d^{2}}{dx^{2}} \ln
(\psi_{0}^{(1)} \cdots \psi _{0}^{(m-1)}), \end{array}
\ea
such that, by knowing all the eigenfunctions and eigenvalues of $H_{1}$ we also obtain the
corresponding eigenfunctions $\psi _{n}^{1}$ and energy eigenvalues $E_{n}^{1}$ of the $(p-1)$ Hamiltonians $(H_{2} \, ,\, \, H_{3} \, ,\, ...,\, H_{p})$.

%

\begin{thebibliography}{99}
\bibitem{Greiner} W. Greiner,
    \href{https://doi.org/10.1007/978-3-662-04275-5}{{\it Relativistics Quantum Mechanics} (third edition, Berlin: Springer), 2000}.
\bibitem{Bagrov} V. G. Bagrov and D. M. Gitman,
    \href{https://www.springer.com/gp/book/9780792302155}{{\it Exact Solutions of Relativistic Wave
Equations} (Dordrecht: Kluwer Academic Publishers), 1990}.
\bibitem{Gara} S. L. Garavelli and F. A. Oliveira,
    \href{https://doi.org/10.1103/PhysRevLett.66.1310}{{\it Phys. Rev. Lett.} {\bf 66}, 1310 (1991)}.
\bibitem{Boivin} L. Boivin, F. X. Kartner and H. A. Haus,
    \href{https://doi.org/10.1103/PhysRevLett.73.240}{{\it Phys. Rev. Lett.} {\bf 73}, 240 (1994)}.
\bibitem{Iwo} I. Bialynicki-Birula, 
    \href{https://link.aps.org/doi/10.1103/PhysRevLett.93.020402}{{\it Phys. Rev. Lett.} {\bf 93}, 020402 (2004)}.
\bibitem{Mili}  M. Belic, N. Petrovic, W. P. Zhong, R. H. Xie and G. Chen, 
    \href{https://doi.org/10.1103/PhysRevLett.101.123904}{{\it Phys. Rev. Lett.} {\bf 101}, 123904 (2008)}.
\bibitem{Herman} H. Feshbach and F. Villars, 
    \href{https://doi.org/10.1103/RevModPhys.30.24}{{\it Rev. Mod. Phys.} {\bf 30}, 24 (1958)}.
\bibitem{KFG1} O. Klein, 
    \href{https://doi.org/10.1007/BF01397481}{{\it Z. Phys.} {\bf 37}, 895 (1926)}.
\bibitem{KFG2} V. Fock,
    \href{https://doi.org/10.1007/BF01399113}{{\it Z. Phys.} {\bf 38}, 242 (1926)}.
\bibitem{KFG3} V. Fock,
    \href{https://doi.org/10.1007/BF01321989}{{\it Z. Phys.} {\bf 39}, 226 (1926)}.
\bibitem{KFG4} W. Gordon,
    \href{http://dx.doi.org/10.1007/BF01390840}{{\it Z. Phys.} {\bf 40}, 117 (1926)}.
\bibitem{Nikiforov} A. F. Nikiforov and V. B.  Uvarov,
    \href{http://dx.doi.org/10.1007/978-1-4757-1595-8}{{\it Special Functions of Mathematical Physics} (Birkh\"{a}user: Basel), 1988}.
\bibitem{Cooper1} F. Cooper, A. Khare and U. Sukhatme,
    \href{https://doi.org/10.1142/4687}{{\it Supersymmetry in Quantum Mechnics} (World Scientific), 2001}.
\bibitem{Cooper2} F. Cooper, A. Khare and U. Sukhatme, 
    \href{https://doi.org/10.1016/0370-1573(94)00080-M}{{\it Phys. Rep.} {\bf 251}, 267-385 (1995)}.
\bibitem{Morales} D. A. Morales, 
    \href{https://doi.org/10.1016/j.cplett.2004.06.109}{{\it Chem. Phys. Lett.} {\bf 394}, 68 (2004)}.
\bibitem{Tang} A. Z. Tang and F. T. Chan, 
    \href{https://doi.org/10.1103/PhysRevA.35.911}{{\it Phys. Rev.} A \textbf{35}, 911 (1987)}.
\bibitem{Roy} B. Roy and R. Roychoudhury, 
    \href{https://doi.org/10.1088/0305-4470/20/10/048}{{\it J. Phys. A: Math. Gen.} \textbf{20}, 3051 (1987)}.
\bibitem{aim} H. Ciftci, R. L. Hall and N. Saad, 
    \href{https://doi.org/10.1088/0305-4470/36/47/008}{{\it J. Phys. A: Math. Gen.} \textbf{36}, 11807 (2003)}.
\bibitem{Hartree} J.C. Slater, 
    \href{https://doi.org/10.1103/PhysRev.81.385}{{\it Phys. Rev.} {\bf 81}, 385 (1951)}.
\bibitem{Cai} J. M. Cai, P. Y. Cai and A. Inomata, 
    \href{https://doi.org/10.1103/PhysRevA.34.4621}{{\it Phys. Rev.} A {\bf 34}, 4621 (1986)}.
\bibitem{Dong1} S. H. Dong,
    \href{https://doi.org/10.1007/978-1-4020-5796-0}{{\it Factorization Method in Quantum Mechanics} (Dordrecht: Springer), 2007}.
\bibitem{Stevenson} P. M. Stevenson, 
    \href{https://doi.org/10.1103/PhysRevD.23.2916}{{\it Phys. Rev.} D {\bf 23}, 2916 (1981)}.
\bibitem{Dong3} G. F. Wei, Z. Z. Zhen and S. H. Dong,
    \href{https://doi.org/10.2478/s11534-008-0143-9}{{\it Cent. Eur. J. Phys.} \textbf{7}, 175 (2009)}.
\bibitem{Jia2013} C. S. Jia, T. Chen, and S. He, 
    \href{https://doi.org/10.1016/j.physleta.2013.01.016}{{\it Phys. Lett.} A \textbf{377}, 682 (2013)}.
\bibitem{Wei2010} G. F. Wei and S. H. Dong,
    \href{https://doi.org/10.1016/j.physletb.2010.02.070}{{\it Phys. Lett.} B. {\bf 686}, 288-292 (2010)}.
\bibitem{Sever2011} A. Arda and R. Sever,
    \href{https://doi.org/10.1063/1.3641246}{{\it  J. Math. Phys.} \textbf{52}, 092101 (2011)}.
\bibitem{Hamzavi2013}  M. Hamzavi, S. M. Ikhdair and  K. E. Thylwe,
    \href{https://doi.org/10.1088/1674-1056/22/4/040301}{{\it  Chin. Phys.} B {\bf 22}, 040301 (2013)}.
\bibitem{Wang2015} Z. Wang, Z. W. Long, C. Y. Long and L. Z. Wang,
    \href{https://doi.org/10.1007/s12648-015-0677-9}{{\it  Indian J Phys.} {\bf 89}, 1059 (2015)}.
\bibitem{Znojil}  M. Znojil, 
    \href{https://doi.org/10.1088/0305-4470/14/2/015}{{\it J. Phys. A: Math. Gen.} \textbf{14}, 383 (1981)}.
\bibitem{Yuan}  C. Y. Chen, D. S. Sun and F. L. Lu, 
    \href{https://doi.org/10.1016/j.physleta.2007.05.079}{{\it Phys. Lett.} A \textbf{370}, 219 (2007)}.
\bibitem{Ikot2011} A. N. Ikot,  L. E. Akpabio and  E. J. Uwah,
    \href{http://ejtp.com/articles/ejtpv8i25p225.pdf}{{\it Electron. J. Theor. Phys.} {\bf 8}, 225 (2011)}.
\bibitem{Mehmet}  M. Simsek and  H. Egrifes,
    \href{https://doi.org/10.1088/0305-4470/37/15/007}{{\it J. Phys. A, Math. Gen.} \textbf{37}, 4379 (2004)}.
\bibitem{Sever} H. Egrifes and R. Sever, 
    \href{https://doi.org/10.1007/s10773-006-9251-8}{{\it Int. J. Theoret. Phys.} \textbf{46}, 935 (2007)}.
\bibitem{Qiang} W. C. Qiang, R. S. Zhou, and Y. Gao, 
    \href{https://doi.org/10.1016/j.physleta.2007.04.109}{{\it Phys. Lett.} A \textbf{371}, 201 (2007)}.
\bibitem{Qiang2004} W. C. Qiang, 
    \href{https://doi.org/10.1088/1009-1963/13/5/002}{{\it Chin. Phys.} {\bf 13}, 575-578 (2004)}.
\bibitem{Guo2005}  J. Y. Guo and  Z. Q. Sheng, 
    \href{https://doi.org/10.1016/j.physleta.2005.02.026}{{\it Phys. Lett.} A. {\bf 338}, 90 (2005)}.
\bibitem{Berkdemir}  C. Berkdemir,  A. Berkdemir,  R. Sever, 
    \href{https://doi.org/10.1088/0305-4470/39/43/005}{{\it J. Phys. A: Math. Gen.} {\bf 39}, 13455 (2006)}.
\bibitem{Badalov} V. H. Badalov,  H. I. Ahmadov and S. V. Badalov, 
    \href{https://doi.org/10.1142/S0218301310015862}{{\it Int. J. Mod. Phys. E} {\bf 19}, 1463 (2010)}.
\bibitem{Oluwadre}  O. J. Oluwadare, K. J. Oyewumi and O. A. Babalola,
    \href{http://lamp.ictp.it/index.php/aphysrev/article/view/543}{{\it Afr. Rev. Phys.} \textbf{7}, 0016 (2012)}.
\bibitem{Ahmadov1}  A. I. Ahmadov, M. Naem, M. V. Qocayeva and V. A. Tarverdiyeva,
    \href{https://doi.org/10.1142/S0217751X18500215}{{\it Int. J. Mod. Phys. A} \textbf{33}, 1850021 (2018)}.
\bibitem{Ahmadov2} A. I. Ahmadov,  Sh. M. Nagiyev,  M. V. Qocayeva, K. Uzun and V. A. Tarverdiyeva,
    \href{https://doi.org/10.1142/S0217751X18502032}{{\it Int. J. Mod. Phys. A} \textbf{33}, 1850203 (2018)}.
\bibitem{Ahmadov3} A. I. Ahmadov, S. M. Aslanova,  M. Sh. Orujova,  S. V. Badalov and  S. H. Dong, 
    \href{https://doi.org/10.1016/j.physleta.2019.06.043}{{\it Phys. Lett. A} \textbf{383}, 3010 (2019)}.
\bibitem{Ahmadov4} A. I. Ahmadov, M. Demirci and S. M. Aslanova,
    \href{https://doi.org/10.1088/1742-6596/1416/1/012001}{{\it J. Phys.: Conf. Ser.} {\bf 1416}, 012001 (2019)}.
\bibitem{Manning1} M. F. Manning, {\it Phys. Rev.} \textbf{44}, 951 (1933).
\bibitem{Manning2} M. F. Manning and  N. Rosen, {\it Phys. Rev.} \textbf{44}, 953 (1933).
\bibitem{Yukawa}  H. Yukawa,
    \href{https://doi.org/10.11429/ppmsj1919.17.0_48}{{\it Proc. Phys. Math. Soc. Jpn.} \textbf{17}, 48 (1935)}.
\bibitem{Gendenshtein1} L. E. Gendenshtein,
    \href{https://ui.adsabs.harvard.edu/abs/1983JETPL..38..356G}{{\it JETP Lett.} \textbf{38}, 356 (1983)}.
\bibitem{Gendenshtein2} L. E. Gendenshtein and I. V. Krive,
    \href{https://doi.org/10.1070/PU1985v028n08ABEH003882}{{\it Sov. Phys. Usp.} \textbf{28}, 645 (1985)}.
\bibitem{Ita} B. I. Ita,  H. Louis, P. I. Amos, T. O. Magu and N. A. Nzeata-ibe,
    \href{https://doi.org/10.9734/PSIJ/2017/34330}{{\it Int. J. Phys. Sci} \textbf{15}, 1-6 (2017)}.
\bibitem{Louis} H. Louis, B. I. Ita, P. I. Amos, O. U. Akakuru, M. M. Orosun, N. A. Nzeata-Ibe and M. Philip,
    \href{http://www.ijcps.org/admin/php/uploads/630_pdf.pdf}{{\it Int. J. Chem. Phys. Sci.} \textbf{7}, 33 (2018)}.
\bibitem{Greene} R. L. Greene, C. Aldrich, 
    \href{https://doi.org/10.1103/PhysRevA.14.2363}{{\it Phys. Rev.} A {\bf 14}, 2363 (1976)}.
\bibitem{Wen1} W. C. Qiang and S. H. Dong,
    \href{https://doi.org/10.1016/j.physleta.2006.10.091}{{\it Phys. Lett.} A \textbf{363}, 169 (2007)}.
\bibitem{Wei} G. F. Wei and S. H. Dong,
    \href{https://doi.org/10.1016/j.physleta.2008.10.064}{{\it Phys. Lett.} A \textbf{373}, 49 (2008)}.
\bibitem{Dong6} W. C. Qiang and S. H. Dong,
    \href{https://doi.org/10.1088/0031-8949/79/04/045004}{{\it Phys. Scr.} \textbf{79}, 045004 (2009)}.
\bibitem{Abramowitz} M. Abramowitz and I. A. Stegun,
    \href{https://doi.org/10.1119/1.15378}{{\it Handbook of Mathematical Functions with Formulas, Graphs and Mathematical Tables}} (Dover, New York), 1964.
\bibitem{Ikhdair09} S. M. Ikhdair, 
    \href{https://doi.org/10.1140/epja/i2009-10758-9}{{\it Eur. Phys. J.} A {\bf 40}, 143 (2009)}.
\end{thebibliography}

\end{document}